\documentclass[aps,prb,reprint,superscriptaddress]{revtex4-2}

\usepackage{graphicx}

\begin{document}

\title{Quasi-two-dimensional Fermi surfaces of the antiferromagnet U$_2$RhIn$_8$ revealed by de Haas-van Alphen measurements}

\author{D.~Aoki}
\affiliation{IMR, Tohoku University, Ibaraki 311-1313, Japan}

\author{Y.~Homma}
\affiliation{IMR, Tohoku University, Ibaraki 311-1313, Japan}

\author{H.~Harima}
\affiliation{Graduate School of Science, Kobe University, Kobe 657-8501, Japan}

\author{I.~Sheikin}
\email[]{ilya.sheikin@lncmi.cnrs.fr}
\affiliation{Laboratoire National des Champs Magn\'{e}tiques Intenses (LNCMI-EMFL), CNRS, UGA, 38042 Grenoble, France}

\date{\today}

\begin{abstract}
We report temperature-dependent Hall effect and low-temperature de Haas-van Alphen (dHvA) effect measurements of the antiferromagnetic heavy-fermion compound U$_2$RhIn$_8$. Temperature dependence of the Hall resistivity suggests a considerable reduction of the carrier density in the antiferromagnetic phase. The observed angular dependence of the dHvA frequencies suggests the existence of three almost ideally two-dimensional Fermi surfaces one of which is quite large. The measured effective masses range from 2$m_0$ to 14$m_0$ for the field applied along the $c$ axis. Local density approximation band-structure calculations performed for the paramagnetic ground state reveal more three-dimensional Fermi surfaces than those observed in the experiment. On the other hand, Fermi surfaces obtained for the antiferromagnetic ground state by band folding are more two dimensional. These calculations account reasonably well for the experimental results assuming a slight modification of the calculated Fermi surfaces.
\end{abstract}

\maketitle

\section{Introduction}

Uranium-based heavy-fermion materials exhibit a variety of exotic ground states including coexistence of ferromagnetism and superconductivity in UGe$_2$~\cite{Saxena2000}, URhGe~\cite{Aoki2001}, UCoGe~\cite{Huy2007}, the enigmatic ``hidden order" state in URu$_2$Si$_2$~\cite{Mydosh2020}, and recently discovered spin-triplet superconductivity in UTe$_2$~\cite{Ran2019}. Electronic structure and, in particular, the Fermi surface (FS) dimensionality are among important ingredients to understand such unusual physical properties. For instance, recent theoretical studies suggest that the presence of a three-dimensional (3D) FS pocket in the electronic structure of UTe$_2$ is a preferable condition for the realization of topological superconductivity in this material~\cite{Ishizuka2019}. However, several de Haas-van Alphen (dHvA) experiments~\cite{Aoki2022,Aoki2023,Eaton2024} reveal only corrugated two-dimensional (2D) FS sheets.

In Ce-based heavy-fermion materials, the possibility to vary the FS dimensionality is realized in the family of Ce$_nT_m$In$_{3n+2m}$ ($T =$ transition metal, $n =$ 1, 2, and $m =$ 0, 1, 2) systems containing a sequence of $n$ CeIn$_3$ conducting layers intercalated by $m$ $T$In$_2$ insulating layers along the $c$ axis. Therefore, within this series, the FS is expected to become more and more 2D with increasing the distance between CeIn$_3$ layers, i. e., with increasing the ratio $m/n$. The first member of the family, CeIn$_3$ ($n = 1$, $m = 0$), crystallizes into a cubic structure and has an isotropic, almost spherical, FS~\cite{Ebihara1993}. The layered structures with $m \neq$ 0 are characterized by strongly anisotropic properties and quasi-2D FSs. Indeed, dHvA measurements revealed quasi-2D FS sheets in monolayer ($n = 1$, $m = 1$) systems CeCoIn$_5$~\cite{Settai2001,Hall2001}, CeIrIn$_5$~\cite{Haga2001}, and CeRhIn$_5$~\cite{Shishido2002,Hall2002,Mishra2021}. The FSs of the bilayer ($n = 2$, $m = 1$) compounds Ce$_2$RhIn$_8$~\cite{Ueda2004,Jiang2015}, Ce$_2$PdIn$_8$~\cite{Goetze2015}, and Ce$_2$PtIn$_8$~\cite{Klotz2018} are also quasi-2D, but the degree of two dimensionality in these materials is smaller than in their monolayer counterparts. Finally, the FS of CePt$_2$In$_7$ ($n = 1$, $m = 2$), in which CeIn$_3$ layers are separated by two PtIn$_2$ layers, was found to consist of almost ideally cylindrical sheets~\cite{Goetze2017} and, thus, to be the most 2D among the so far discovered members of the family.

The isostructural U-based family is much less well studied. So far, it contains only three known materials, UIn$_3$, URhIn$_5$, and U$_2$RhIn$_8$. All three are antiferromagnets with rather high N\'{e}el temperatures. Similar to CeIn$_3$, the FS of UIn$_3$ is 3D and isotropic~\cite{Tokiwa2001}. In URhIn$_5$, dHvA measurements revealed four oscillatory frequencies, two of which seem to originate from large cylindrical FSs, while the other two correspond to small 3D pockets~\cite{Yu2017}. To the best of our knowledge, the electronic structure of U$_2$RhIn$_8$ has not been reported so far.

U$_2$RhIn$_8$ crystalizes into a tetragonal crystal structure with space group $P4/mmm$ (no. 123)~\cite{Bartha2015} shown in the inset of Fig.~\ref{Rho&Cp}. It undergoes an antiferromagnetic (AF) transition at a N\'{e}el temperature $T_N =$ 117 K. The N\'{e}el temperature does not change upon application of the magnetic field up 9~T along the $c$ axis. However, $T_N$ monotonically increases as a function of pressure up to 3.2~GPa. The Sommerfeld coefficient $\gamma =$~47~mJ/K$^2$mol is rather large given the high N\'{e}el temperature. The effective magnetic moment, $\mu_{eff} = 3.45 \mu_B/U$ extracted from the temperature dependence of magnetic susceptibility, is close to both the U$^{4+}$ (5$f^2$) and U$^{3+}$ (5$f^3$) moments of 3.58$\mu_B/U$ and 3.63$\mu_B/U$, respectively~\cite{Tokiwa2001a}. The magnetic structure determined by single-crystal neutron diffraction is commensurate with propagation vector $\mathbf{Q} = (1/2, 1/2, 0)$ and magnetic moments aligned along the $c$ axis~\cite{Bartha2019}. Interestingly, Ce$_2$RhIn$_8$ exhibits the same propagation vector~\cite{Bao2001}.

In this paper, we report Hall and dHvA effect measurements in U$_2$RhIn$_8$. The observed FS is dominated by three almost cylindrical sheets. The effective masses are found to be moderately enhanced, up to 14$m_0$, where $m_0$ is the bare electron mass. The experimental results are compared with band-structure calculations performed for both paramagnetic (PM) and AF ground states.

\section{Experimental details}

\begin{figure}[htb]
\includegraphics[width=\columnwidth]{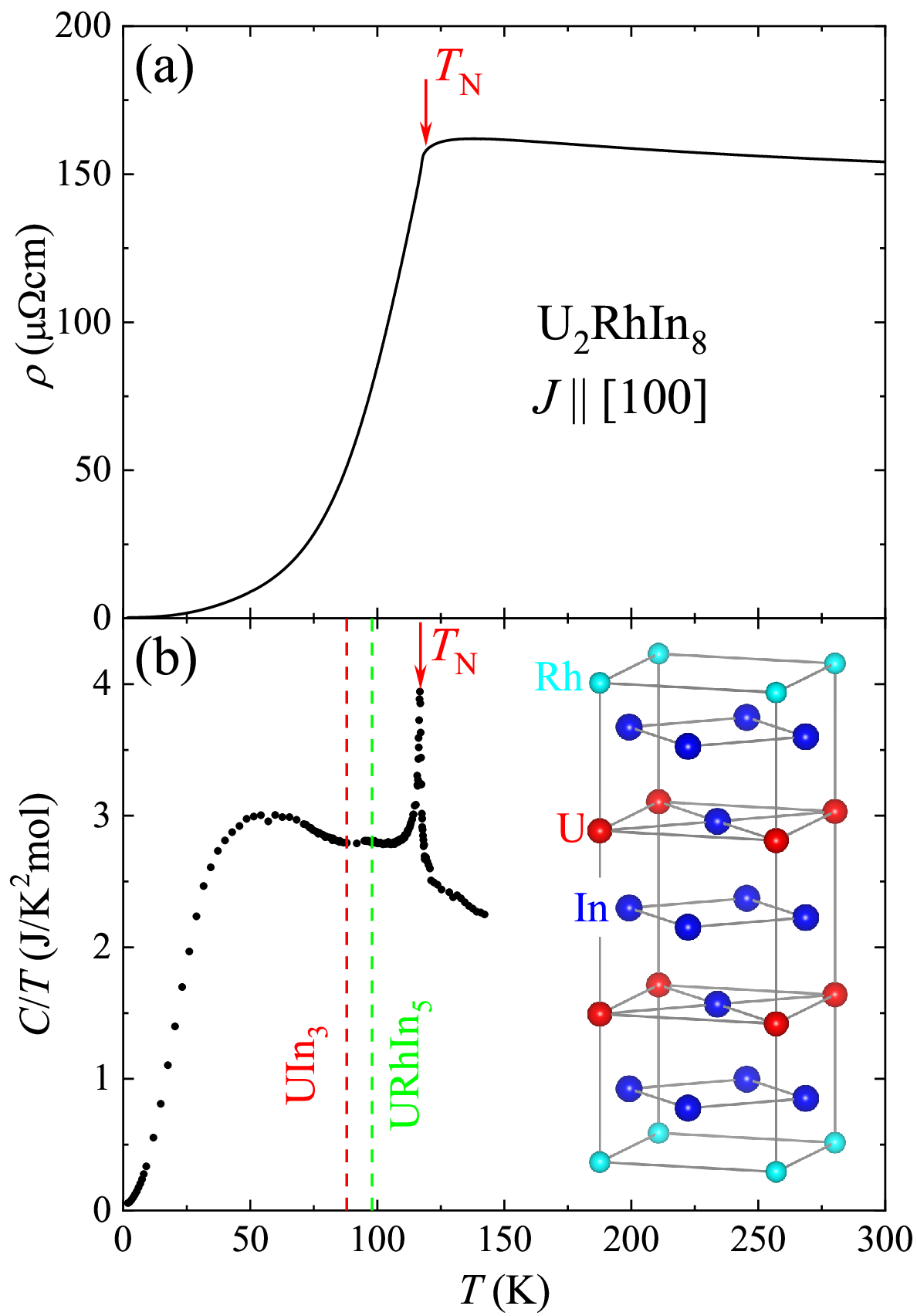}
\caption{\label{Rho&Cp} (a) Temperature dependence of the electrical resistivity in U$_2$RhIn$_8$ for current applied along the $a$ axis. (b) Specific heat divided by temperature as a function of temperature. Arrows indicate the anomalies corresponding to the AF transition at $T_N =$~117~K. Dashed lines correspond to the N\'{e}el temperatures of URhIn$_5$ and UIn$_3$. The inset shows the crystal structure of U$_2$RhIn$_8$.}
\end{figure}

Single crystals of U$_2$RhIn$_8$ used for this study were grown by an In self-flux method, similar to that described in Ref.~\cite{Bartha2015}, using high-purity elements U(99.95\%), Rh(99.9\%), and In(99.9999\%). The starting composition of U:Rh:In = 2:1:25 was placed in an alumina crucible, which was then sealed in an evacuated quartz tube. The ampoule was heated up to 950 $^\circ$C in 8 h, kept at this temperature for 12 h, and then cooled down to 600 $^\circ$C in 122 h. The excess In was removed in a centrifuge. The obtained single crystals were plate-like with typical dimensions of $1 \times 1 \times 0.3$~mm$^3$.

The crystal structure of the grown crystals was confirmed using a single-crystal x-ray diffractometer (Rigaku XtaLAB mini II) with graphite monochromated Mo $K_\alpha$ radiation ($\lambda = 0.71075$~{\AA}). Of the 2247 Bragg reflections collected, 348 were unique. The crystal structure was solved with SHELXT and then refined with SHELXL. The atomic coordinates and the equivalent isotropic atomic displacement parameters $B_{eq}$ of U$_2$RhIn$_8$ are shown in Table~\ref{atomic_positions}. The lattice constants $a$ and $c$ of the tetragonal crystal structure are determined to be 4.6133(3)~{\AA} and
12.0275(12)~{\AA}, respectively. These parameters are similar to those previously reported~\cite{Bartha2015}. The single crystals were oriented using a Laue camera (Photonic Science Laue x-ray CCD camera).

\begin{table}
\caption{\label{atomic_positions}Atomic coordinates and equivalent isotropic atomic displacement parameters $B_{eq}$ of U$_2$RhIn$_8$ at 23 $^\circ$C. The number in parentheses is the standard deviation of the last digits. The least-squares refinement was based on 348 independent reflections and converged with a conventional agreement factor $R_1 =$~0.0435.}
\begin{ruledtabular}
\begin{tabular}{lccccc}
Atom& Site& $x$& $y$& $z$& $B_{eq}$\\
\hline
U& 2$g$& 0& 0& 0.30865(7)& 0.51(3)\\
Rh& 1$a$& 0& 0& 0& 0.60(5)\\
In(1)& 2$e$& 0& 1/2& 1/2& 0.54(4)\\
In(2)& 2$h$& 1/2& 1/2& 0.30898(13)& 0.46(4)\\
In(3)& 4$i$& 0& 1/2& 0.12225(11)& 0.62(3)\\
\end{tabular}
\end{ruledtabular}
\end{table}

In order to assess the crystal quality, we performed temperature-dependent electrical resistivity measurements. The measurements were performed using a conventional ac four-point technique down to 1.7~K in a quantum design physical property measurement system (QD PPMS). The same technique was used for Hall resistivity measurements. The latter measurements were carried out with six point contacts using a bar-shaped sample with a thickness of 0.09 mm and a width of 0.74 mm. In order to avoid the asymmetry of the contacts, temperature dependence was measured both in positive and negative field of 1~T, and the magnetoresistance component was subtracted. Linear field dependence of the transverse resistance was observed at 1.7 K at least up to 1T. The resistivity curve for current applied along the $a$ axis is shown in Fig.~\ref{Rho&Cp}(a). A clear kink at 117~K indicated by arrow manifests the AF transition. No other anomalies were observed down to 1.7~K. The residual resistivity ratio (RRR) and the residual resistivity, $\rho_0$, were determined to be 700 and 0.22~$\mu \Omega$cm, respectively, indicating very high quality of our crystals.

It is well known that some Ce-based 218 crystals sometimes contain a small amount of the CeIn$_3$ phase~\cite{Kratochvilova2014,Klotz2018}. Furthermore, previous work on U$_2$RhIn$_8$ indicated that at least some of the single crystals grown by In-self flux technique contained a small amount of URhIn$_5$~\cite{Bartha2015}, while some of the previously grown URhIn$_5$ crystals were contaminated with a trace amount of U$_2$RhIn$_8$~\cite{Yu2017}. In order to verify the purity of our single crystals, we measured specific heat using QD PPMS. Temperature dependence of specific heat divided by temperature, $C/T$, is shown in Fig.~\ref{Rho&Cp}(b). A distinct $\lambda$-shaped anomaly at 117~K indicates the second-order AF transition. No other clear anomalies were observed down to 1.8~K, the lowest temperature of our measurements. In particular, no well defined anomalies typical for an AF transition were observed at either 98~K or 88~K, the N\'{e}el temperatures of URhIn$_5$ and UIn$_3$, respectively. This implies that our crystals do not contain any detectable amount of impurity phases.

dHvA effect measurements were performed by standard field-modulation technique in a top-loading dilution refrigerator equipped with a low-temperature rotator. The measurements were performed in magnetic fields up to 14.7~T and temperatures down to 60 mK.

\section{Results and discussion}

\subsection{Hall effect}

\begin{figure}[htb]
\includegraphics[width=\columnwidth]{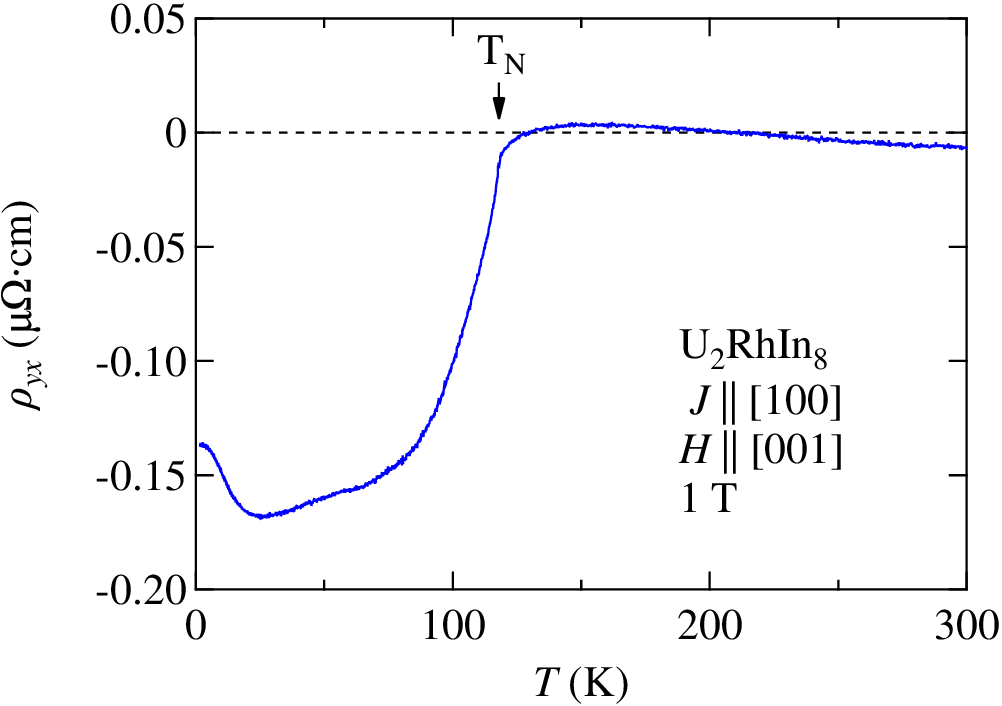}
\caption{\label{Hall} Temperature dependence of the Hall resistivity in U$_2$RhIn$_8$ for current applied along the $a$ axis and magnetic field of 1~T applied along the $c$ axis.}
\end{figure}

Figure~\ref{Hall} shows temperature dependence of the Hall resistivity for the field along the $c$ axis measured at 1~T and for the current along the $a$ axis. The Hall resistivity shows a weak temperature dependence on cooling from 300~K and then strongly decreases below $T_N$. The Hall resistivity consists of two components - that is normal and anomalous Hall resistivity. The former one is inversely proportional to the carrier number and the latter is proportional to the magnetization. Given that the magnetization is strongly temperature dependent above $T_N$~\cite{Bartha2015}, weak temperature dependence with the values close to zero at a high temperature region above $T_N$ indicates that the contribution from the anomalous part of the Hall resistivity is small. The normal Hall resistivity is also small because of the large carrier number due to the fact that U$_2$RhIn$_8$ is an uncompensated metal in the PM state. In the AF state with the propagation vector $\mathbf{Q} = (1/2, 1/2, 0)$, the unit cell is doubled. Therefore, the electronic state changes into a compensated metal, in which the total carrier numbers from electrons and holes could be reduced. In fact, as shown in Fig.~\ref{Hall}, the Hall resistivity shows a large negative value revealing the reduction of carrier numbers.

\subsection{dHvA oscillations}

\begin{figure}[htb]
\includegraphics[width=\columnwidth]{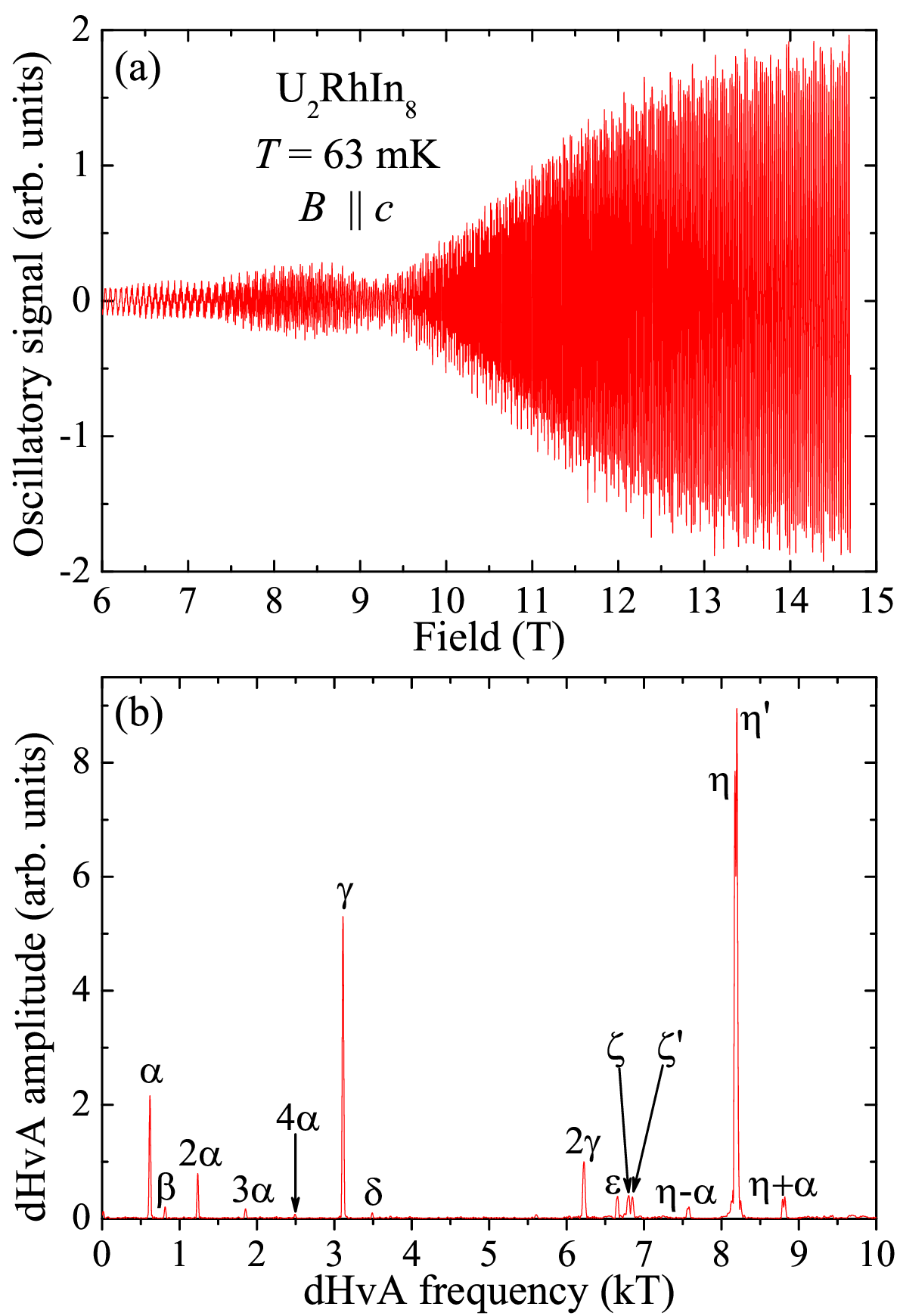}
\caption{\label{Oscillations} (a) Oscillatory signal after subtracting the nonoscillating background in U$_2$RhIn$_8$ for magnetic field applied along the $c$ axis at 63 mK. (b) FFT spectrum of the dHvA oscillations from (a) over the field range from 6 to 14.7~T.}
\end{figure}

Figure~\ref{Oscillations}(a) shows the oscillatory signal after subtracting the nonoscillating background for the magnetic field applied along the $c$ axis at 63~mK. The low-frequency oscillations are clearly visible at a field as low as 6~T. This confirms the high quality of our crystals. At higher fields, high-frequency oscillations become dominant.

The fast Fourier transform (FFT) spectrum  of the dHvA oscillations from Fig.~\ref{Oscillations}(a) is shown in Fig.~\ref{Oscillations}(b). For the field applied along the $c$ axis, we observed seven fundamental frequencies denoted as $\alpha$, $\beta$, $\gamma$, $\delta$, $\epsilon$, $\zeta$, and $\eta$. In addition, we observed multiple harmonics of some of the dHvA frequencies as well as some linear combinations, e. g., $\eta+\alpha$ and $\eta-\alpha$. Two fundamental frequencies, $\zeta$ and $\eta$, seem to be split into two close satellites. Details of the angular dependence of the
dHvA frequencies will be discussed later on, together with the results of the band-structure calculations.

\subsection{Band-structure calculations}

\begin{figure}[htb]
\includegraphics[width=\columnwidth]{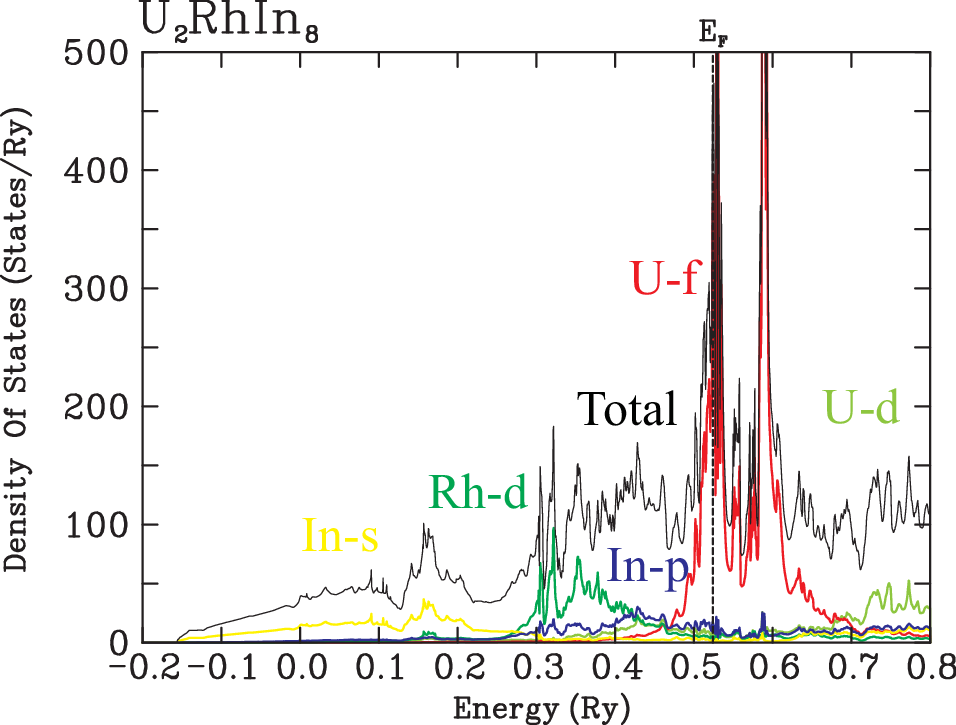}
\caption{\label{DOS} Calculated density of states for U$_2$RhIn$_8$. The dashed line indicates the Fermi level $E_F$.}
\end{figure}

Band-structure calculations assuming the PM ground state of U$_2$RhIn$_8$ were performed using the KANSAI code, which provides a full potential linear augmented plane wave method with the local density approximation (LDA), where the spin-orbit coupling is considered as a second variational procedure~\cite{Harima1990}. The band structure and FSs in the actual AF ground state were obtained through the band-folding of the PM bands at the boundaries of the magnetic Brillouin zone (BZ) with $\mathbf{Q} = (1/2, 1/2, 0)$, as discussed in more detail elsewhere~\cite{Kubo1993}.

\begin{figure}[htb]
\includegraphics[width=\columnwidth]{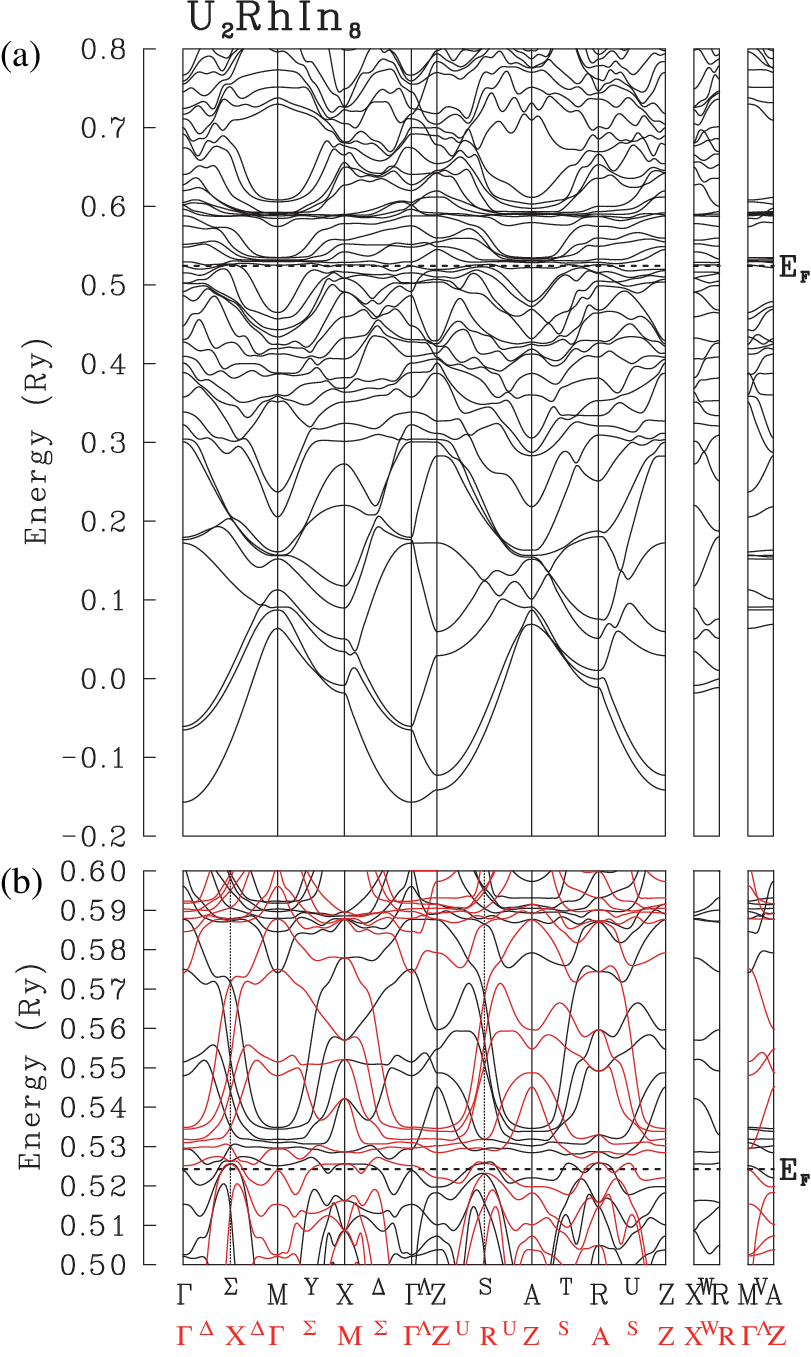}
\caption{\label{Band_structure} (a) Calculated band structure along high-symmetry axes for the PM ground state of U$_2$RhIn$_8$. (b) Zoom of the band structure in the vicinity of the Fermi level for PM (black) and AF (red) ground states. The latter was obtained by folding of the PM bands, as explained in the text. The dashed lines indicate the Fermi level $E_F$.}
\end{figure}

The resulting density of states, which is not affected by the band folding, and band structures are shown in Figs.~\ref{DOS} and~\ref{Band_structure}, respectively.

Remarkably, the calculated density of states at the Fermi level, $E_F$, is high exceeding 450~states/Ry. For comparison, the calculated density of states at $E_F$ is at least several times smaller in other U-based compounds, such as UPd$_2$Al$_3$~\cite{Inada1999}, UGa$_2$~\cite{Honma2000}, UPd$_3$~\cite{Tokiwa2001b}, and UPt$_5$~\cite{Sato2020}. The electronic specific heat coefficient obtained from the calculated density of states at $E_F$ is $\gamma_b =$~40~mJ/K$^2$mol. This value is very close to the experimental one, $\gamma =$~47~mJ/K$^2$mol~\cite{Bartha2015}. This is again an exception, as calculated Sommerfeld coefficients are usually a few times smaller than those measured by specific heat in U-based compounds.

\begin{figure*}[htb]
\includegraphics[width=\textwidth]{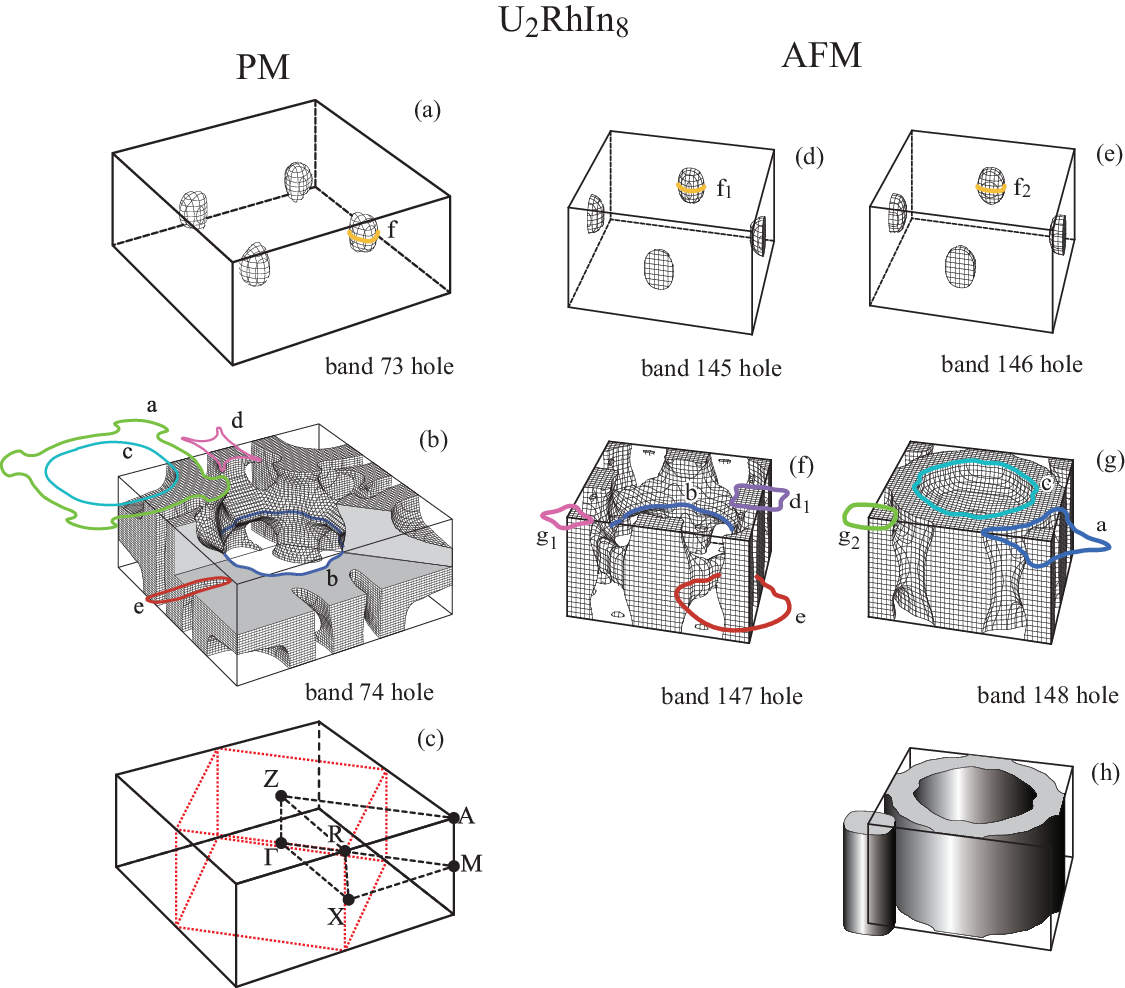}
\caption{\label{Fermi_surface} Panels (a) and (b) calculated FSs for U$_2$RhIn$_8$ in the PM ground state. Solid lines indicate extremal cross sections for the field applied along the $c$ axis. (c) PM Brillouin zone with high symmetry points. Dotted lines indicate the magnetic BZ. Panels (d) - (g) calculated FSs for U$_2$RhIn$_8$ in the AF ground state. Solid lines indicate extremal cross sections for the field applied along the $c$ axis. (h) Schematic illustration of the 2D FSs, which might originate from band 148 in the AF ground state, as discussed in the text.}
\end{figure*}

The calculated FSs in the PM ground state are shown in Figs.~\ref{Fermi_surface}(a) and \ref{Fermi_surface}(b). The FS originating from band 73 consists of small hole pockets. The FSs originating from band 74 are much larger. They include ellipsoidal electron pockets centered at the $A$ point and a more complex multiconnected hollow hole sheet, which occupy approximately 50\% of the BZ, as expected for an uncompensated metal.

As was already discussed above, AF ordering modifies the BZ. This, in turn, leads to the modification of the FSs due to band folding of the PM bands. In the particular case of U$_2$RhIn$_8$, the band folding affects most of the FSs of the PM ground state, as can be seen in Figs.~\ref{Fermi_surface}(d) - \ref{Fermi_surface}(g). Folding of the PM band 73 results in small ellipsoidal hole pockets in bands 145 and 146. The FSs of the AF bands 147 and 148 are the result of the folding of the PM band 74. These FSs are more 2D than their PM counterparts. In particular, there are corrugated cylindrical FSs of band 147 in the corners of the magnetic BZ. The other hollow sheet of the FS originating from band 147 also contains quasi-2D parts along the $X-R$ direction of the magnetic BZ. Similarly, there are quasi-2D parts of the large FS of band 148 in the corners of the magnetic BZ. The large ellipsoidal electron sheet centered at the $Z$ point of the magnetic BZ is essentially the same as its PM counterpart originating from the PM band 74.

\subsection{Angle dependence of the dHvA frequencies and comparison with band-structure calculations}

\begin{figure*}[htb]
\includegraphics[width=\textwidth]{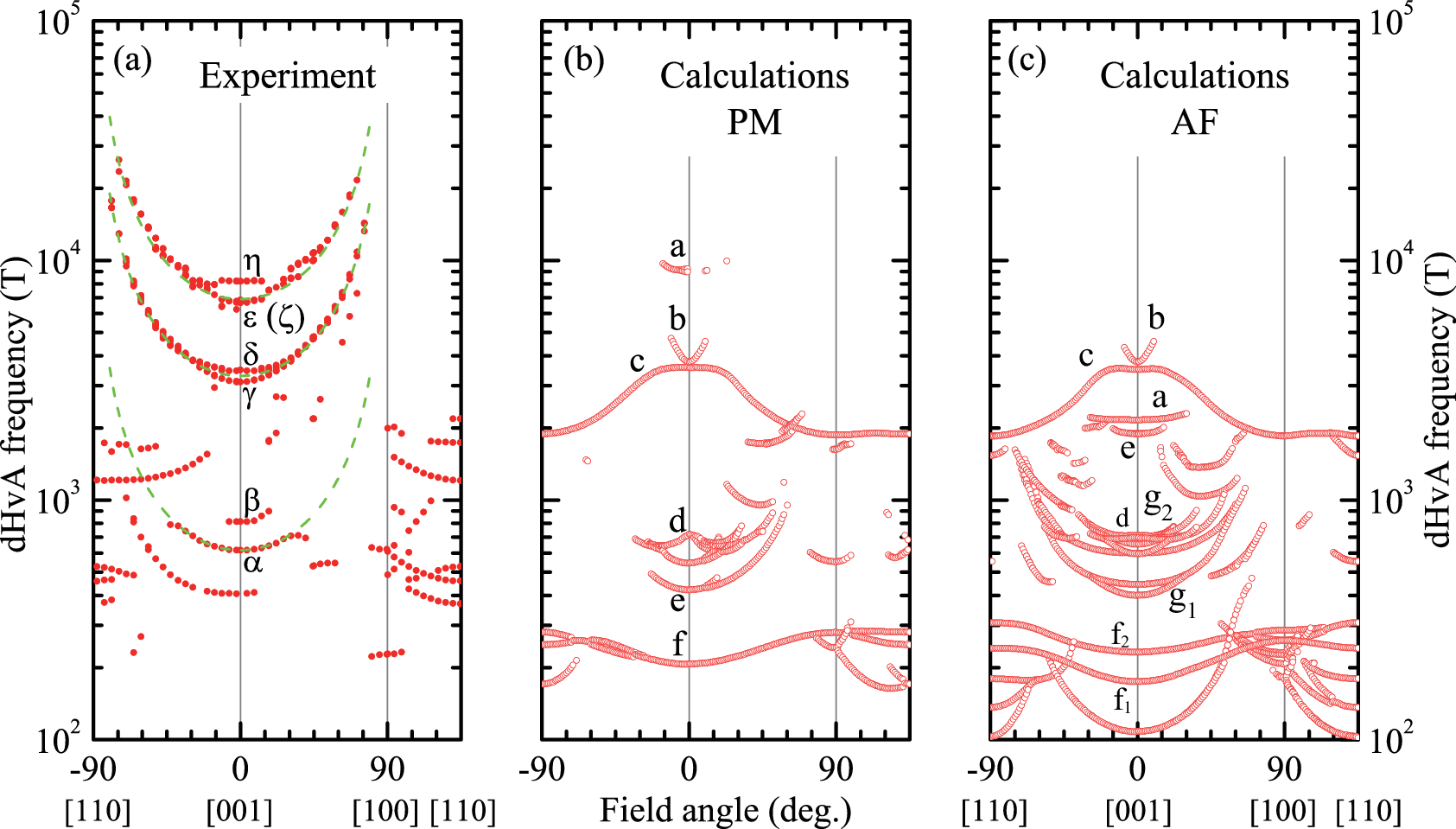}
\caption{\label{Comparison} Angle dependence of the experimentally observed (a) and calculated dHvA frequencies in PM (b) and AF (c) ground states in U$_2$RhIn$_8$. Dashed lines represent $1 / \cos (\theta)$ dependencies characteristic of cylindrical FSs.}
\end{figure*}

The angle dependence of the dHvA frequencies is shown in Fig.~\ref{Comparison}(a). Branches $\alpha$, $\gamma (\delta)$, and $\varepsilon$ almost ideally follow the $1/\cos (\theta)$ dependence, where $\theta$ is the angle from the $c$ axis. This dependence, shown by the dashed line in Fig.~\ref{Comparison}(a), is characteristic for a cylindrical FS found in 2D systems. Therefore, these branches originate from three distinct almost cylindrical FS sheets. The $\alpha$ branch is observed up to 35\---40$^\circ$ from the $c$ axis, while $\gamma (\delta)$ and $\varepsilon$ branches can be traced up to 70\---80$^\circ$. The two branches $\gamma$ and $\delta$ are close in frequency and converge at a small angle from the $c$ axis. This suggests that they arise on the same quasi-2D FS sheet. The same is true for the $\varepsilon$ and $\eta$ branches. The $\beta$ branch is observed only close to the $c$ axis. It is not clear whether it corresponds to yet another quasi-2D sheet or the same one as the $\alpha$ branch. Several branches observed around the [110] direction probably correspond to more complex multiconnected FSs. Remarkably, we have not observed any branches, which extend over the whole angle range. This rules out the existence of 3D isotropic FS pockets such as spherical or ellipsoidal.

The FSs in U$_2$RhIn$_8$ appear to be more 2D than those reported in the sister compound URhIn$_5$~\cite{Yu2017}. This is rather surprising given that the situation is the opposite in Ce-based compounds~\cite{Goetze2015}.

For U$_2$RhIn$_8$, the cross section of the PM BZ perpendicular to the $c$ axis is $1.85 \times 10^{20}$~m$^{-2}$. The AF order, however, modifies the BZ. For U$_2$RhIn$_8$ in the AF state, the propagation vector $\mathbf{Q} = (1/2, 1/2, 0)$. Therefore, the cross section of the AF BZ is two times smaller than its PM counterpart, $9.27 \times 10^{19}$~m$^{-2}$. It still exceeds the largest FS cross section $S_{\mathrm{max}} = 7.83 \times 10^{19}$~m$^{-2}$ corresponding to the highest dHvA frequency, 8.2~kT, of the $\eta$ branch. Therefore, all the observed FS sheets fit into the AF BZ.

Calculated angle dependence of the dHvA frequencies in the PM state shown in Fig.~\ref{Comparison}(b) reveals a certain degree of similarity with the experimental results. For example, the calculations predict two branches, $e$ and $d$, with upward curvature around the $c$ axis with frequencies about 400 T and 550 T, respectively, for fields along the $c$ axis. One of them might correspond to the experimentally observed $\alpha$ branch. Furthermore, the calculations reveal another high-frequency, $F \simeq 9$~kT, $a$ branch around the $c$ axis, which also shows a positive curvature. This branch might correspond to the experimentally observed $\varepsilon$ branch, although the calculated branch exists over a small angular range only. However, the experimentally observed branches $\gamma$  and $\delta$ originating from the same cylindrical sheet of the FS are not reproduced. Furthermore, the calculated small hole pockets from band 73 corresponding to the $f$ branch and ellipsoidal electron sheets from band 74 centered at the $A$ point giving rise to the $c$ branch are not observed in the experiment. Overall, the experimentally observed FSs appear to be considerably more 2D than those revealed by the band-structure calculations in the PM ground state.

Interestingly, similar band-structure calculations reveal quasi-2D FS sheets in Ce$_2$PdIn$_8$~\cite{Goetze2015} and Ce$_2$PtIn$_8$~\cite{Klotz2018}. Furthermore, for these compounds, the agreement between the calculations and the results of the dHvA measurements was found to be very good. On the contrary, LDA calculations performed for the PM ground state of URhIn$_5$ do not agree with the experimental results obtained in this compound~\cite{Yu2017}.

We will now compare experimentally observed dHvA frequencies with those calculated for the AF ground state shown in Fig.~\ref{Comparison}(c). The experimentally observed $\alpha$ branch can be easily accounted for by the calculated branches originating from quasi-2D sheets of bands 147 and 148 located in the corners of the magnetic BZ. At first glance, nothing in the calculations corresponds to the experimentally observed $\gamma$, $\delta$, $\varepsilon$ and $\eta$ branches. Let us, however, have a closer look at the large hole FS from band 148. The main part of this FS connected along $\Gamma-X$ and $\Gamma-M$ directions is otherwise almost ideally cylindrical. If the connections are removed, the maximum and minimum cross sections of this 2D sheet would readily account for the experimentally observed $\eta$ and $\varepsilon$ branches, respectively. Furthermore, breaking the connection along the $\Gamma-M$ direction would create truly 2D sheets in the corners of the magnetic BZ. The maximum cross section of such a FS can then account for the experimental $\beta$ branch. In a similar way, if the electron FSs originating from the same band are cylindrical rather than ellipsoidal, then, assuming a small degree of corrugation, the maximum and minimum cross sections would nicely account for experimental $\delta$ and $\gamma$ branches, respectively. This scenario is schematically illustrated in Fig.~\ref{Fermi_surface}(h).

\subsection{Effective masses}

\begin{figure}[htb]
\includegraphics[width=\columnwidth]{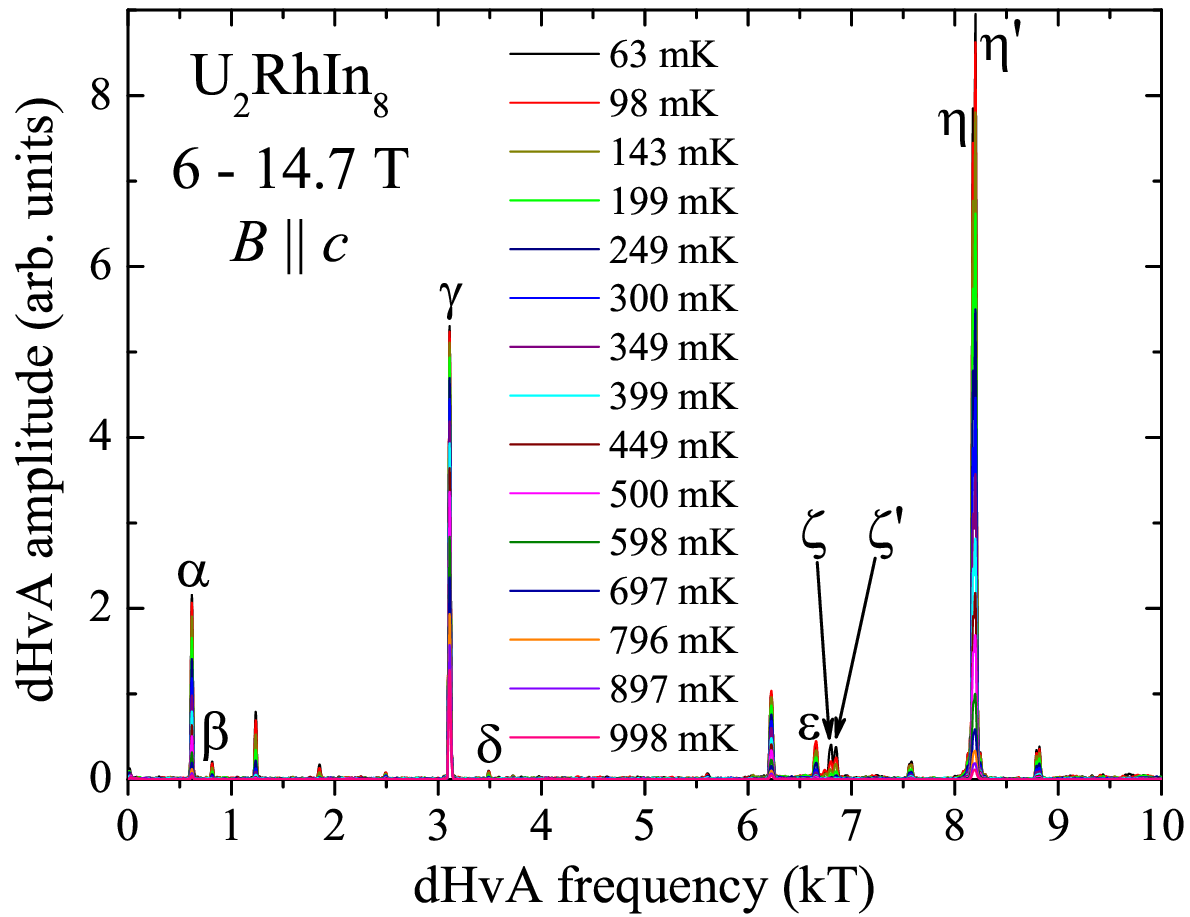}
\caption{\label{T_dep} FFT spectra of the dHvA oscillations measured at different constant temperatures for the field applied along the $c$ axis. Only fundamental frequencies are labeled.}
\end{figure}

To obtain the effective masses, we have measured dHvA oscillations at several constant temperatures. The FFT spectra of the oscillations measured at different temperatures for the field applied along the $c$ axis are shown in Fig.~\ref{T_dep}. The effective masses were determined by fitting the temperature dependence of the oscillatory amplitude by the standard Lifshitz-Kosevich formula~\cite{Shoenberg1984}, as shown in Fig. 9 for the same field orientation. The dHvA frequencies and corresponding effective masses for the field applied along three high-symmetry directions, [001], [100], and [110] are shown in Table~\ref{dHvA parameters}.

\begin{figure}[htb]
\includegraphics[width=\columnwidth]{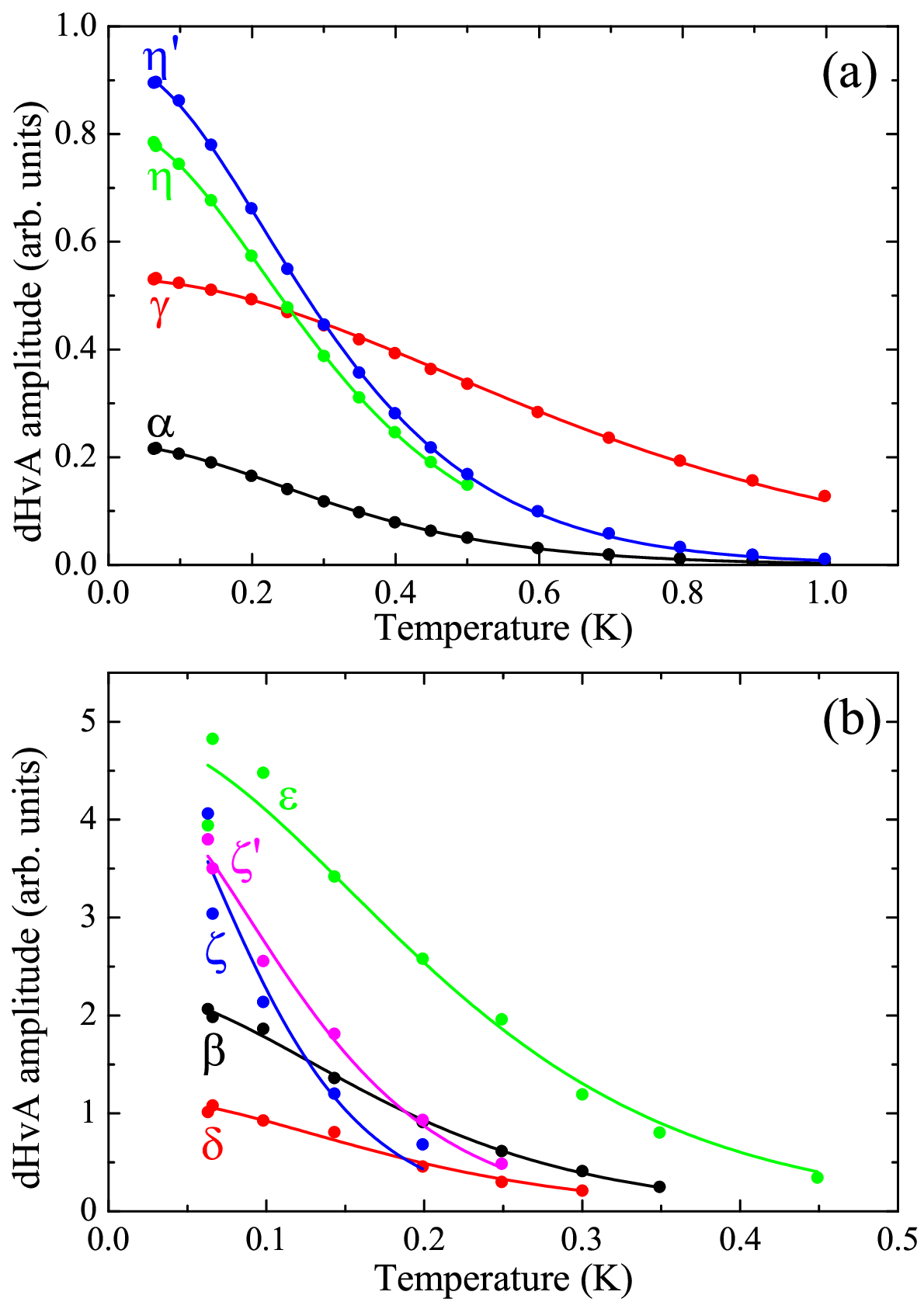}
\caption{\label{Eff_mass} Amplitude of the strong (a) and weak (b) dHvA oscillations in U$_2$RhIn$_8$ as a function of temperature for the field applied along the $c$ axis. The lines are fits by the standard Lifshitz-Kosevich formula~\cite{Shoenberg1984}.}
\end{figure}

For the magnetic field applied along the $c$ axis, the effective masses are moderately enhanced ranging from 2$m_0$ ($m_0$ is the bare electron mass) for the $\gamma$ branch to 14$m_0$ for the $\zeta$ branch. These masses are somewhat higher than those measured in URhIn$_5$, where they range from 1.9$m_0$ to 4.3$m_0$~\cite{Yu2017}. On the other hand, considerably higher masses ranging from 10$m_0$ to 33$m_0$ were measured in UIn$_3$~\cite{Tokiwa2001}. This is rather surprising given that the electronic specific heat coefficients are similar in all three compounds: $\gamma =$~40~mJ/K$^2$mol for UIn$_3$~\cite{Tokiwa2001}, $\gamma =$~50~mJ/K$^2$mol for URhIn$_5$~\cite{Matsumoto2013}, and $\gamma =$~47~mJ/K$^2$mol for U$_2$RhIn$_8$~\cite{Bartha2015}.

\begin{table}[htb]
\caption{\label{dHvA parameters}Experimentally determined dHvA frequencies $F$ and effective masses $m^*$ in U$_2$RhIn$_8$ for the magnetic field applied along three high-symmetry directions. $m_0$ is the bare electron mass. Branch assignments refer to Fig.~\ref{Oscillations}.}
\begin{ruledtabular}
\begin{tabular}{lccc}
Field direction& Branch& $F$ (kT)& $m^{\ast}/m_0$\\
\hline
$B \parallel [001]$& $\alpha$& 0.62& 3.97$\pm$0.02\\
& $\beta$& 0.81& 7.3$\pm$0.1\\
& $\gamma$& 3.11& 1.98$\pm$0.01\\
& $\delta$& 3.49& 7.2$\pm$0.3\\
& $\varepsilon$& 6.66& 6.1$\pm$0.4\\
& $\zeta$& 6.80& 14$\pm$2\\
& $\zeta^{\prime}$& 6.85& 10.5$\pm$0.5\\
& $\eta$& 8.18& 4.34$\pm$0.01\\
& $\eta^{\prime}$& 8.20& 4.34$\pm$0.01\\
\hline
$B \parallel [100]$& & 0.23& 3.24$\pm$0.05\\
& & 0.49& 5.38$\pm$0.06\\
& & 0.61& $<$0.5\\
& & 2.00& 7.64$\pm$0.06\\
\hline
$B \parallel [110]$& & 0.37& 3.37$\pm$0.03\\
& & 0.46& $<$0.5\\
& & 0.53& 5.29$\pm$0.08\\
& & 1.21& 4.54$\pm$0.02\\
& & 1.74& 7.2$\pm$0.2\\
& & 2.19& $<$0.5\\
\end{tabular}
\end{ruledtabular}
\end{table}

For the magnetic field applied in the basal plane, i.e., $B \parallel [100]$ and $B \parallel [110]$, the effective masses are somewhat lower than their counterparts measured along the [001]. Furthermore, the effective masses could not be determined for some of the dHvA frequencies as the corresponding dHvA amplitudes do not change over the temperature range of the measurements, $0.06 \: \mathrm{K} \lesssim T \lesssim 0.7 \: \mathrm{K}$. This implies that these masses do not exceed 0.5$m_0$. Single crystals grown by an In self-flux method sometimes contain a small amount of the remaining In flux. This indium flux gives rise to strong dHvA oscillations characterized by very low effective masses. This scenario, although not impossible, is unlikely in our case. Indeed, the remaining In flux, if present, is in a polycrystalline form. Therefore, dHvA oscillations originating from In flux are expected to be observed for all the orientations of the magnetic field. In our measurements, however, no oscillations with very low masses were observed for the field along the $c$ axis.

Knowledge of the effective masses allows us to estimate the Sommerfeld coefficient. For a cylindrical FS, the  Sommerfeld coefficient is given by $\gamma = k_\mathrm{B}^2 V / (6 \hbar^2) m^* k_z$~\cite{Aoki2000}, where $V$ is the molar mass and $k_z$ is the height of the BZ. Our measurements reveal the presence of three cylindrical sheets of the FS - small ($\alpha$ branch), medium ($\gamma$ and $\delta$ branches), and large ($\varepsilon$, $\zeta$ and $\zeta^{\prime}$ branches). Taking the average effective masses, the contributions from these cylinders are $\gamma_s =$~8.3~mJ/K$^2$mol, $\gamma_m =$~9.6~mJ/K$^2$mol and $\gamma_l =$~21.4~mJ/K$^2$mol, respectively. This sums up to 39.3~mJ/K$^2$mol if we assume one copy of each sheet per magnetic BZ. This value is comparable to the experimental one, $\gamma =$~47~mJ/K$^2$mol. The missing $\gamma$ value might be accounted for by the contributions from other more 3D and/or multiconnected FSs. Alternatively, there might be multiple copies of some of the sheets. For example, if we assume the presence of two copies of the small cylinder, this would account, quite precisely, for all of the specific heat.

\section{Conclusions}

In summary, we performed dHvA and Hall effect measurements in a high-quality single crystal of U$_2$RhIn$_8$. A considerable reduction of the carrier density at the AF transition is suggested by the temperature dependence of the Hall resistivity. The observed FSs are dominated by three almost ideally cylindrical sheets, as expected from the layered structure of the material. The total contribution of these sheets to the electronic specific heat coefficient accounts for at least 80\% of the experimental value. On the other hand, we have not observed any isotropic clearly 3D FS sheets, such as spherical or ellipsoidal. The measured effective masses are found to be moderately enhanced, up to 14$m_0$. LDA band-structure calculations assuming all-itinerant 5$f$ electrons performed for the PM ground state reveal major discrepancies with the experimentally observed dHvA frequencies and their angle dependence. However, the FSs in the AF ground state obtained by a simple band-folding procedure appear to be more 2D and agree rather well with the experimental results assuming a slight modification of the calculated FSs. Such a modification might arise due to effects, which are not taken taken into account in the present calculations. Therefore, more sophisticated calculations are desirable to reliably reproduce the FSs in U$_2$RhIn$_5$.

Theoretically, reduced dimensionality of the FS is expected to enhance unconventional superconductivity~\cite{Monthoux2003}, which is often observed in heavy-fermion systems in the vicinity of a quantum critical point. Such a tendency was indeed observed experimentally in the family of Ce$_nT_m$In$_{3n+2m}$ ($T$: transition metal, $n =$ 1, 2, $m =$ 0, 1, 2) systems. The superconducting critical temperature of cubic CeIn$_3$ with isotropic FS is very low, $T_c \approx$ 0.2~K~\cite{Mathur1998}, even in the vicinity of its critical pressure. This is to be compared with $T_c \approx$ 2.1~K in pressure-induced superconductors CeRhIn$_5$~\cite{Hegger2000} and CePt$_2$In$_7$~\cite{Bauer2010} with quasi-2D FSs. In CeCoIn$_5$, which also exhibits quasi-2D FSs, superconductivity occurs at ambient pressure at $T_c =$ 2.3~K~\cite{Petrovic2001}. This is the highest value observed so far in Ce-based heavy-fermion compounds. U$_2$RhIn$_8$ is not superconducting; neither are the other two known compounds of its family, UIn$_3$ and URhIn$_5$. However, superconductivity with critical temperatures comparable to or even exceeding those in Ce-based compounds was observed in some U-based materials, such as, e.g., U$_6$Fe ($T_c \approx$ 4~K) and U$_6$Co ($T_c \approx$ 2.3~K)~\cite{Chandrasekhar1958}. The electronic structure of theses materials is not known. However, angular dependence of the upper critical field suggests rather isotropic ellipsoidal FS~\cite{Yamamoto1996,Aoki2016}. Given almost ideally 2D FSs observed in U$_2$RhIn$_8$, it would be interesting to grow 218 U-based compounds with other transition metals, such as Co, Ir, Pt or Pd. If some of them turn out to be superconducting, the transition temperature might be relatively high.

\begin{acknowledgments}
This work was supported by the ICC-IMR of Tohoku University and JSPS KAKENHI Grants No. JP24K00587, No. JP19H00646, No. JP20K20889, No. JP20H00130, No. JP20KK0061, No. JP22H04933, and No. JP24H01641.
\end{acknowledgments}

\bibliography{U2RhIn8_dHvA}

\end{document}